\documentstyle[12pt]{article}
\author {Wen-Xiu Ma\footnote{Institute of Mathematics, Fudan 
University, 
Shanghai 200433, P. R. of China} $\,$\footnote{Mathematik und 
Informatik, 
Universit\"at-GH Paderborn, D-33098 Paderborn, Germany} $\ $
 and Fu-Kui Guo\footnote{Dept. of Applied Mathematics, Shandong Mining
Institute, Tai'an 271019, P. R. of China}}
\title
{Lax Representations and Zero Curvature Representations by Kronecker 
Product}
\date{\nonumber}
\begin{document}
\maketitle

\setlength{\baselineskip}{19pt}
\def \be {\begin{equation}}
\def \ee {\end{equation}}
\def \ba {\begin{array}}
\def \ea {\end{array}}
\def \bea {\begin{eqnarray}}
\def \eea {\end{eqnarray}}
\def \la {\lambda}
\def \al {\alpha}

\begin{abstract}
It is showed that Kronecker product can be applied to construct 
not only new Lax representations but also new
 zero curvature representations of integrable 
models. Meantime a different characteristic between continuous and 
discrete 
zero curvature equations is pointed out. 
\end{abstract}

\newtheorem{thm}{Theorem}
\newtheorem{Le}{Lemma}

Lax representation and zero curvature representation play an 
important role
in studying nonlinear integrable models in theoretical physics.
 It is based on
such representations that the inverse scattering transform
is successfully developed (see, say, Ablowitz and Clarkson 1991). 
They may also provide a lot of 
information, such as integrals of motion, master
symmetries and Hamiltonian formulation.
There exist quite many integrable models to possess Lax 
representation or
 zero curvature representation (Faddeev and Takhtajan 1987, Das 
1989). 
Two typical examples are Toda lattice (Flaschka 1974)  and 
AKNS systems (Ablowitz, Kaup, Newell and Segur 1974) including KdV 
equation
and nonlinear Sch\"odinger equation. 

In this paper, we want to give rise to a kind of new Lax 
representations 
and new zero curvature representations by using Kronecker product of 
matrices,
motivated by  a recent progress made by Steeb and Heng (Steeb and 
Heng 1996).
Kronecker product itself has nice mathematical properties 
and important applications in many fields of physics, for example,
statistical physics, 
quantum groups, etc. (Steeb 1991). Our result for 
zero curvature representation also provides us with a different 
characteristic
between continuous and discrete zero curvature equations.

Let $I_M$ denote the unit matrix of order $M, \ M\in Z$. For two 
matrices 
$A=(a_{ij})_{pq},\,B=(b_{kl})_{rs}$, Kronecker product
$A\otimes B$ is defined by (Steeb 1991)
\be A\otimes B = (a_{ij}B)_{(pr)\times (qs)},\ee 
or equivalently by (Hoppe 1992)
\be (A\otimes B)_{ij,kl}=a_{ik}b_{jl}.\ee 
 Evidently we have a basic relation on Kronecker product (Steeb 1991,
Hoppe 1992)
\be (A\otimes B)(C\otimes D)=(AC)\otimes (BD),\label{basic}\ee 
provided that the matrices $AC$ and $BD$ make sense. This relation 
will be 
used to show new structure of Lax representation and zero 
curvature representation of integrable models.  

\begin{thm} {\rm (Lax representation)}
Assume that an integrable model (continuous or discrete) has two Lax 
representations
\be L_{1t}=[A_1,L_1],\ L_{2t}=[A_2,L_2],\label{twolr}\ee
where $L_1,\,A_1$ and $L_2,\,A_2$ are $M\times M$ and $N\times N$ 
matrices,
respectively. Define 
\be L_3=\al _1 L_1\otimes L_2+\al _2 (L_1\otimes I_N+I_M\otimes L_2), 
\ 
A_3=A_1\otimes I_N+I_M\otimes A_2, \ee
where $\al _1,\, \al _2$ are arbitrary constants.
Then the same integrable model has another Lax representation 
$ L_{3t}=[A_3,L_3].$
\end{thm}
{\bf Proof:} 
First of all, we have 
\be L_{3t}= \al _1 (L_{1t}\otimes L_2+L_1\otimes L_{2t})
+\al _2 (L_{1t}\otimes I_N+I_M\otimes L_{2t}).\nonumber \ee 
On the other hand, using (\ref{basic}) we can calculate that
\bea 
&& [A_{3},L_{3}]=\al _1([A_1,L_1]\otimes L_2+L_1\otimes[A_2,L_2])
\nonumber \\ &&\qquad \qquad
+\al _2([A_1,L_1]\otimes I_N+I_M\otimes [A_2,L_2]).\nonumber
\eea 
Now we easily find that the equalities defined by 
(\ref{twolr}) implies $L_{3t}=[A_3,L_3]$.
$\vrule width 1mm height 3mm depth 0.4mm$

When $\al _2 =0$, the obtained result is exactly one in Ref. 
Steeb and Heng 1996. When $\al _1 =0$, we get a new Lax 
representation  for 
a given integrable model, starting from two known Lax 
representations. 
Integrals of motion may also 
be generated from new Lax representation, because we have
\bea F_{ij}&=&\textrm{tr}(\al _1 L_1^i\otimes L_2^j+\al 
_2(L_1^i\otimes I_N+I_M
\otimes L_2^j))\nonumber \\ &
=&\al _1 \textrm{tr}(L_1^i)\textrm{tr}(L_2^j)+\al _2
(N\textrm{tr}(L_1^i)+M\textrm{tr}(L_2^j)),\label{img} \eea 
where we have used $ \textrm{tr}(A\otimes 
B)=\textrm{tr}(A)\textrm{tr}(B)$
(Steeb 1991) and $(L_1^i)_t=[A_1,L_1^i],\ (L_2^i)_t=[A_2,L_2^i]$.

\begin{thm} {\rm (Continuous zero curvature representation)} 
Assume that a continuous integrable model has two continuous zero 
curvature
representations
\be U_{1t}-V_{1x}+[U_1,V_1]=0,\ U_{2t}-
V_{2x}+[U_2,V_2]=0,\label{twoczcr}\ee
where $U_1,\,V_1$ and $U_2,\,V_2$ are $M\times M$ and $N\times N$ 
matrices,
respectively. Define 
\be U_3=U_1\otimes I_N+I_M\otimes U_2, \ 
V_3=V_1\otimes I_N+I_M\otimes V_2. \ee
Then the same integrable model has another continuous zero curvature 
representation  
\be U_{3t}-V_{3x}+[U_3,V_3]=0.\label{thirdcr}\ee 
\end{thm}
{\bf Proof:} The proof is also a direct computation.
We first have 
\bea && U_{3t}= U_{1t}\otimes I_N+I_M\otimes U_{2t},\nonumber \\ &&
U_{3x}= U_{1x}\otimes I_N+I_M\otimes U_{2x}.\nonumber \eea
Second, using (\ref{basic}) we can obtain that
\be 
[U_{3},V_{3}]=[U_1,V_1]\otimes I_N+I_M\otimes [U_2,V_2].\nonumber
\ee 
Therefore we see that (\ref{thirdcr}) is true once two equalities 
defined by (\ref{twoczcr}) hold.
$\vrule width 1mm height 3mm depth 0.4mm$

We remark that when we choose 
\[ U_3=U_1\otimes U_2,\]
the third zero curvature representation  (\ref{thirdcr}) is not 
certain to 
be true. An example will be displayed later on.

\begin{thm}
{\rm (Discrete zero curvature representation)} 
Assume that a discrete integrable model has two discrete zero 
curvature
representations
\be U_{1t}=(EV_1)U_1-U_1V_1,\ U_{2t}=(EV_2)U_2-
U_2V_2,\label{twodzcr}\ee
where $E$ is the shift operator,  
$U_1,\,V_1$ are $M\times M$ matrices, and $U_2,\,V_2$ are $N\times N$ 
matrices. Define
\be U_3=U_1\otimes U_2, \ 
V_3=V_1\otimes I_N+I_M\otimes V_2. \ee
Then the same integrable model has another discrete zero curvature 
representation  
\be U_{3t}=(EV_3)U_3-U_3V_3.\label{thirddr}\ee 
\end{thm}
{\bf Proof:} 
Similarly, we first have 
\be  U_{3t}= U_{1t}\otimes U_2+U_1\otimes U_{2t}.\nonumber \ee 
On the other hand, we may calculate that
\bea && (EV_3)U_3-U_3V_3=((EV_1)\otimes I_N+I_M\otimes 
(EV_2))(U_1\otimes U_2)
\nonumber \\ && \qquad \qquad\quad
-(U_1\otimes U_2)(V_1\otimes I_N+I_M\otimes V_2)\nonumber \\ &&
=((EV_1)U_1)\otimes U_2+U_1\otimes ((EV_2)U_2)-(U_1V_1)
\otimes U_2-U_1\otimes (U_2V_2)\nonumber \\ &&
=((EV_1)U_1-U_1V_1)\otimes U_2+U_1\otimes ((EV_2)U_2-U_2V_2)
.\nonumber
\eea 
In the second equality above, we have used the basic relation 
(\ref{basic}).
Hence we find that (\ref{thirddr}) holds if we have (\ref{twodzcr}).
$\vrule width 1mm height 3mm depth 0.4mm$

We remark that when we choose 
\[ U_3=U_1\otimes I_M+I_N\otimes U_2,\]
the third discrete zero curvature representation  (\ref{thirddr}) is 
not 
certain to be true. An example will also be given later on.
This is opposite to the result in the continuous case. 
It shows us a different characteristic
between continuous and discrete zero curvature equations.

In what follows, we would like to display some concrete examples to 
illustrate
 the use of the above technique of Kronecker product.
Actually once we have a Lax representation or a zero curvature 
representation,
we can obtain a new representation after choosing two required 
representations
 to be this known one. Further newer 
representation may be constructed 
by use of this new representation and the process 
may be infinitely proceeded to.
This also tells us that there exist 
infinitely many Lax representations or zero curvature representations
once there exists one representation for a given integrable model.
The concrete procedure of construction will
be showed in the following examples and can be easily generalized to 
other 
integrable models, for example, in Refs. Calogero 
1994, Drinfel'd and Sokolov 1984,
 Ma 1993, Ragnisco and Santini 1990, Tu 1990 etc.

\noindent {\bf Example 1:} We consider periodical
Toda lattice (Flaschka 1974)
\be a_{it}=a_i(b_{i+1}-b_i),\ b_{it}=2(a_i^2-a_{i-1}^2),\ 
a_{i+N}=a_i,\
 b_{i+N}=b_i,\label{tl}\ee
which is a Hamiltonian system with Hamiltonian
\[H(q_1,q_2,\cdots, q_N,p_1,p_2,\cdots,p_N)=\frac12 \sum_{i=1}^N 
p_i^2+
\sum_{i=1}^N\textrm{e}^{q_i-q_{i+1}}\]
under the Flaschka's transformation
\[a_i=\frac 12\textrm{e}^{q_i-q_{i+1}},\ b_i=-\frac12 p_i. \] 
Toda lattice (\ref{tl}) has a Lax representation  with 
\bea && L= \left (\ba {cccccc} b_1& a_1 &0&\cdots &\cdots & 
a_N\vspace{2mm} \\ 
a_1&b_2&a_2&\cdots &\cdots &0 \vspace{2mm} \\ 
0&a_2&b_3&\cdots &\cdots &0\vspace{2mm} \\ 
\vdots & &\ddots &\ddots & &\vdots\vspace{2mm} \\ 
\vdots & & &\ddots &\ddots &a_{N-1}\vspace{2mm} \\ 
a_N&\cdots &\cdots &\cdots &a_{N-1}&b_N  \ea \right),\\ && 
A= \left (\ba {cccccc} 0& a_1 &0&\cdots &\cdots & -a_N\vspace{2mm} \\ 
-a_1&0&a_2&\cdots &\cdots &0 \vspace{2mm} \\ 
0&-a_2&0&\cdots &\cdots &0\vspace{2mm} \\ 
\vdots & &\ddots &\ddots & &\vdots\vspace{2mm} \\ 
\vdots & & &\ddots &\ddots &a_{N-1}\vspace{2mm} \\ 
a_N&\cdots &\cdots &\cdots &-a_{N-1}&0\ea \right). \eea 
Through Theorem 1, we obtain a new Lax representation  with
 \be L_{{new}}=\al _1  L\otimes L+\al _2( L\otimes I_N+I_N\otimes L)
,\ A_{{new}}= A\otimes I_N+I_N\otimes A.\ee
Here $\al _1,\,\al _2 $ are two arbitrary constants and thus Toda 
lattice
(\ref{tl})
has a lot of different Lax representations. By (\ref{img}), new 
integrals 
of motion may be generated, which are all functions of 
$F_i=\textrm{tr}(L^i)$.

\noindent {\bf Example 2:}
Nonlinear Sch\"odinger model (Ablowitz, Kaup, Newell and Segur 1974, 
Ma and Strampp 1994)
\be \left \{\ba {l} p_t=-\frac12 q_{xx}+p^2q,\vspace{2mm}\\
q_t=\frac12 p_{xx}-pq^2 \ea \right.\label{NLSS}\ee 
has a continuous zero curvature representation  with 
\be U= \left (\ba {cc} -\la & p \vspace{2mm} \\ q &  \la   \ea 
\right),\ 
V= \left (\ba {cc} -\la ^2+\frac12 pq& \la p-\frac12 p_x\vspace{2mm} 
\\
 \la q +\frac1 2 q_x & \la ^2 -\frac12 pq  \ea \right). \ee 
This model has infinitely many symmetries and integrals of motion.
According to 
 Theorem 2, we obtain new continuous zero curvature representations 
 with
\bea  U_{{new}}&=& U\otimes I_2+I_2\otimes U= \left (\ba {cccc}
-2\la &p &p &0 \vspace{2mm}\\ 
q& 0& 0& p\vspace{2mm}\\ q& 0& 0& p\vspace{2mm}\\ 
0&q&q&2\la 
 \ea\right), \label{newso}\vspace{2mm} \\ 
V_{{new}}&= &V\otimes I_2+I_2\otimes V= \left (\ba {cccc}
-2\la ^2 +pq &\la p -\frac12 p_x&\la p -\frac12 p_x&0 \vspace{2mm}\\ 
\la q+\frac12 q_x& 0& 0& \la p -\frac12 p_x\vspace{2mm}\\
\la q+\frac12 q_x& 0& 0& \la p -\frac12 p_x\vspace{2mm}\\
0&\la q+\frac12 q_x &\la q+\frac12 q_x& 2\la ^2 -pq 
 \ea\right), \qquad
\eea
or with 
 \be U_{{new}}= U\otimes I_4+I_2\otimes 
\left (\ba {cc} U& 0 \vspace{2mm} \\ 0 & U \ea \right),\ 
V_{{new}}= V\otimes I_4+I_2\otimes\left (\ba {cc} V& 0 \vspace{2mm} 
\\ 0 & V
 \ea \right) . \nonumber\ee 
The above spectral operator defined by (\ref{newso}) is  similar to 
one 
appearing in Khasilev 1992 and we may also discuss its binary 
nonlinearization
(for the cases of $2\times 2$ and $3\times 3$ matrices, see
Ma and Strampp 1994, Ma, Fuchssteiner and Oevel 1996).
However nonlinear Sch\"odinger model (\ref{NLSS})
 haven't the continuous zero curvature representation
with 
 \be U_{{new}}= U\otimes 
  U ,\ 
V_{{new}}= V\otimes I_2+I_2\otimes V .\nonumber\ee 

\noindent {\bf Example 3:} We consider
Bogoyavlensky lattice (Fuchssteiner and Ma 1996) 
\be u_t=u(u^{(-1)}-u^{(1)}),\ u^{(m)}=E^mu .\label{bl}\ee
More examples of lattice may be found in Refs. Steeb 1991,
Ragnisco and Santini 1990
and Tu 1990 etc., for example, one-dimensional isotropic Heisenberg 
model. The lattice (\ref{bl})
has a discrete zero curvature representation  with 
\be U= \left (\ba {cc} 1& u \vspace{2mm} \\ \la ^{-1} &0  \ea 
\right),\ 
V= \left (\ba {cc} \frac12 \la -u& \la u \vspace{2mm} 
\\ 1 & -\frac12 \la -u^{(-1)}
  \ea \right). \ee 
By Theorem 3, we obtain new discrete zero curvature representations 
with
\be U_{{new}}= U\otimes \left (\ba {cc} U& 0 \vspace{2mm} 
\\ 0 & U \ea \right),\ 
V_{{new}}= V\otimes  I_4+I_2\otimes \left( \ba {cc} V& 0 \vspace{2mm} 
\\ 0 & V \ea \right),\nonumber \ee
or with  
 \be U_{{new}}= \left (\ba {cc} U& 0 \vspace{2mm} 
\\ 0 & U \ea \right)\otimes 
\left (\ba {ccc} U& 0 &0\vspace{2mm} \\ 0 & U &0\vspace{2mm} 
\\ 0&0&U \ea \right),\ 
V_{{new}}= \left( \ba {cc} V& 0 \vspace{2mm} 
\\ 0 & V \ea \right)\otimes 
I_6 +I_4\otimes\left( \ba {ccc} V& 0 &0\vspace{2mm} \\ 
0&V&0\vspace{2mm} 
\\ 0&0 & V
 \ea \right). \nonumber\ee 
The latter is two $24\times 24$ matrices. 
If we want to directly find these two matrices, we will meet 
a lot of complicated calculation. It is worth to point out that 
we haven't the discrete zero curvature representation
with 
 \be U_{{new}}= U\otimes I_2+I_2\otimes
  U ,\ 
V_{{new}}= V\otimes I_2+I_2\otimes V \nonumber\ee 
for Bogoyavlensky lattice (\ref{bl}).
This is not strange and shows us a difference between two kinds of 
zero 
curvature representations.

Finally we present an open problem. We denote the Gateaux derivative 
$K'[S]$
by $ K'[S]=\frac {\partial }{\partial \varepsilon }K'(u+\varepsilon 
S)|_
{\varepsilon =0}.$
We have already established the following result
(Ma 1992, Ma 1993, Fuchssteiner and Ma 1996):
If $u_t=K(u),\ u_t=S(u)$ have Lax representations 
\[ L_t=[A_1,L],\ L_t=[A_2, L]\] 
or zero curvature representations
\[\ba{l}
 U_{t}-V_{1x}+[U,V_1]=0\ \ (\,\textrm{or}\ U_t=(EV_1)U-
UV_1\,),\vspace{1mm}\\
U_{t}-V_{2x}+[U,V_2]=0\ \ (\,\textrm{or}\ U_t=(EV_2)U-UV_2\,),\ea\]
respectively, then the product model 
$u_t=[K,S]:=K'[S]-S'[K]$ has Lax representation
\[L_t=[A_3,L],\ A_3=A_1'[S]-A_2'[K]+[A_1,A_2]\]
or zero curvature representation
\[ U_{t}-V_{3x}+[U,V_3]=0\ \ (\,\textrm{or}\ U_t=(EV_3)U-UV_3\,),\ 
V_3=V_1'[S]-V_2'[K]+[V_1,V_2].\]
Therefore $u_t=[K,S]$ has Lax representation with the spectral 
operator 
and the Lax operator determined by Kronecker product.  For example, 
in the case of Lax representation we have
\bea && L_{new}=\al _1 L\otimes L+\al _2 (L\otimes I_M+I_M\otimes L),
\nonumber \\ && A_{new}= (A_1'[S]-A_2'[K]+[A_1,A_2])\otimes 
I_M+I_M\otimes 
(A_1'[S]-A_2'[K]+[A_1,A_2]),
\nonumber
\eea
where $M$ is the order of the matrix $L$.
Product models may be applied to construct symmetries of nonlinear 
models and thus they are important.
Let us now suppose that two models 
$u_t=K(u),\ u_t=S(u)$ have two completely different 
Lax representations \[L_{1t}=[A_1,L_1],\ L_{2t}=[A_2,L_2]\]
 or two zero curvature representations 
\[ \ba {l}U_{1t}-V_{1x}+[U_1,V_1]=0\ \ (\,\textrm{or}\ U_t=(EV_1)U-
UV_1\,),
\vspace{1mm}\\
U_{2t}-V_{2x}+[U_2,V_2]=0\ \ (\,\textrm{or}\ U_t=(EV_2)U-UV_2\,).\ea 
\]
 Here $L_1$ and $L_2$ or $U_1$ and $U_2$ are not equal, and sometimes
they may have  
different orders of matrix.
The problem is what the corresponding 
representation for the product
model $u_t=[K,S]$ is.
It seems to us that the required spectral operator
matrix $L_{new}$ or $U_{new}$ should be represented by
some Kronecker product involving $L_1,L_2$ or $U_1,U_2$ and $K,S$.

\newpage \small 
{\bf REFERENCES}
\vskip 2mm  
\begin{description}
\item Ablowitz, M. J. and Clarkson, P. A. (1991). {\it Solitons, 
Nonlinear
Evolution Equations and Inverse Scattering}, Cambridge University 
Press, 
Cambridge.
\item Ablowitz, M. J., Kaup, D. J., Newell, A. C. and Segur, H. 
(1974).
{\it Studies in Applied Mathematics}, {\bf 53}, 249. 
\item Calogero, F. and Nucci, M. C. (1991). {\it Journal of 
Mathematical
 Physics}, {\bf 32}, 72.
\item Das, A. (1989). {\it Integrable Models}, {World Scientific, 
Singapore}.
\item Drinfel'd, V. G. and Sokolov, V. V. (1984). {\it Soviet Journal 
of Mathematics}, {\bf 24}, 81.
\item Faddeev, L. D. and Takhtajan, L. A. (1987). {\it Hamiltonian 
Methods
in the Theory of Solitons}, Springer-Verlag, Berlin.
\item Flaschka, H. (1974). {\it Physical Review B}, {\bf 9}, 1924.
\item
Fuchssteiner, B. and Ma, W. X. (1996). 
Master symmetries of discrete models by discrete zero curvature 
equations,
to appear in Proceedings of Symmetries and Integrability of 
Differential 
Equations, Canterbury, UK.
\item Khasilev, V. Ya. (1992). {\it JETP Letters}, {\bf 56}, 194. 
\item Hoppe, J. (1992). {\it Lectures on Integrable Systems}, 
Springer-Verlag, Berlin.
\item Ma, W. X. (1992). {\it Journal of Mathematical
 Physics}, {\bf 33}, 2464.
\item
Ma, W. X. (1993). {\it Journal of Physics A: Mathematical and 
General},
{\bf 26}, 2573.
\item Ma, W. X. (1993). {\it Journal of Physics A: Mathematical and 
General},
{\bf 26}, L1169.
\item Ma, W. X., Fuchssteiner, B. and Oevel, W. (1996). 
A $3\times 3$ matrix spectral problem for AKNS hierarchy and 
    its binary Nonlinearization, to appear in {\it Physica A}.
\item
Ma, W. X. and Strampp, W. (1994). {\it Physics Letters A}, {\bf 185}, 
277.
\item Ragnisco, O and Santini, P. M. (1990). {\it Inverse Problems}, 
{\bf 6}, 441.
\item
Steeb, W.-H. (1991). {\it Kronecker Product and Applications}, 
BI-Wissenschaftsverlag, Mannheim.
\item
Steeb, W.-H. and Heng, L. C. (1996). {\it International  Journal
of Theoretical Physics}, {\bf 35},
475. 
\item
Tu, G. Z. (1990). {\it Journal of Physics A: Mathematical and 
General},
{\bf 23}, 3903.
\end{description}
\end{document}